\documentclass[]{article}

\usepackage{amssymb,latexsym,amsmath}     
\usepackage{authblk}
\usepackage[utf8]{inputenc}
\usepackage{dsfont}
\usepackage[utf8]{inputenc}
\usepackage{graphicx}
\usepackage{mathtools}
\usepackage{amsmath}
\usepackage{subfig}
\usepackage{xcolor}
\usepackage{epstopdf}
\usepackage{placeins}
\usepackage{empheq}
\usepackage{tikz}
\usepackage{siunitx,array,multirow}
\usepackage[colorlinks = true,
linkcolor = blue,
urlcolor  = blue,
citecolor = blue,
anchorcolor = blue]{hyperref}
\usepackage[numbers,square,sort&compress]{natbib}
\usetikzlibrary{positioning,decorations.pathmorphing}
\usepackage{pifont}
\usepackage{multirow}
\usepackage[width=\textwidth]{caption}

\begin{document}
	\title{Generalized volume-complexity for Lovelock black holes}
	
	\author{Monireh Emami \thanks{monireh.emami@modares.ac.ir}}
	\author{Shahrokh Parvizi \thanks{parvizi@modares.ac.ir}}

	\affil{Department of Physics, School of Sciences,\\
	Tarbiat Modares University,\\
	 P.O.Box 14155-4838, Tehran, Iran}

	\maketitle
	\begin{abstract}
	We study the time dependence of the generalized complexity of Lovelock black holes using the ``complexity = anything" conjecture, which expands upon the notion of ``complexity = volume" and generates a large class of observables. By applying a specific condition, a more limited class can be chosen, whose time growth is equivalent to a conserved momentum. Specifically, we investigate the numerical full time behavior of complexity time rate, focusing on the second and third orders of Lovelock theory coupled with Maxwell term, incorporating an additional term -the square of the Weyl tensor of the background spacetime- into the generalization function. Furthermore, we repeat the analysis for case with three additional scalar terms: the square of Riemann and Ricci tensors, and the Ricci scalar for second-order gravity (Gauss-Bonnet) showing how these terms can affect to multiple asymptotic behavior of time. We study how the phase transition of generalized complexity and its time evolution occur at turning point $(\tau_{turning})$ where the maximal generalized volume supersedes another branch. Additionally, we discuss the late time behavior, focusing on proportionality of the complexity time rate to the difference of temperature times entropy at the two horizons ($TS(r_+)-TS(r_-)$) for charged black holes, which can be corrected by generalization function of each radius in generalized case. In this limit, we also explore near singularity structure by approximating spacetime to Kasner metrics and finding possible values of complexity growth rate with different choices of the generalization function. 
	\end{abstract}
	\section{Introduction}
	
	In recent decades, the AdS/CFT correspondence has emerged as a powerful framework for investigating quantum gravity \cite{mal}. Moreover, quantum information theory has shed light on the theory of quantum gravity through this framework, for example, by quantities such as entanglement entropy and quantum complexity \cite{RT,Van}. Computational complexity, in the context of information theory,  is defined as the minimum number of unitary quantum operators known as gates, that evolve an initial state into a final requiring state \cite{comcom}. This notion has provided tools to study the complexity of CFT the in boundary, and gravitational theory in the bulk. The latter was led to CV, CA, and CV2.0 conjectures.\\
	The CV conjecture, proposed within the AdS/CFT framework, is the first conjecture suggesting this physical quantity, motivated by ER=EPR \cite{ms}, to justify the growth of the Einstein-Rosen Bridge (ERB) ``size" over time in an eternal AdS black hole, after the scrambling time \cite{fast} and the thermal equilibrium of dual boundary theories. Accordingly, the complexity can be quantified by the maximum volume of a hypersurface anchored in the boundary time slice $\Sigma_{CFT}$ (the spatial volume of ERB) where the boundary CFT state resides, expressed as \cite{comcom,cv2}
	\begin{align}\label{cv}
	\mathcal{C}(\Sigma_{CFT})= \max_{\partial\Sigma=\Sigma_{CFT}} \frac{V(\partial\Sigma)}{G_N L},
	\end{align}
	where $L, G_L$ and $\Sigma$ are length scale, gravitational constant and extremal hypersurface anchored in the boundary time slice $\Sigma_{CFT}$, respectively. \\
	In another attempt, CA conjecture was proposed that the complexity can be related to the action of a specific extremal surface in the bulk spacetime, so-called Wheeler-DeWitt (WDW) patch \cite{ca1,ca2}
	\begin{align}\label{ca}
	\mathcal{C}(\Sigma)= \frac{I_{WDW}}{\pi\hbar},
	\end{align} 
	where $I_{WDW}$ is action of Wheeler-DeWitt patch and $\hbar$ is Planck constant. In \cite{cv2.0}, authors define volume complexity for the Wheeler-DeWitt (WDW) patch as subregion complexity motivated by \cite{alish,ben-car}, and propose CV2.0 conjecture of complexity as
	\begin{align}
	\mathcal{C}(\Sigma)=\frac{V_{WDW}}{G_N L^2},
	\end{align}
	where $V_{WDW}$ is volume of Wheeler-DeWitt patch. There are some inherent ambiguities in the definitions of quantum complexity, such as choosing the reference state, adopting the unitary set of operators, and so on. Correspondingly, in the bulk theory, one expects there would be some freedom in introducing the holographic complexity. Recently, ``generalized complexity" (or complexity = anything, CAny) conjecture has emerged as a generalized version of complexity in bulk, where complexity is not limited to a specific geometric quantity but can be described by a more general function \cite{any,any?,anys}. This conjecture allows for a broader exploration of complexity measures beyond volume and action, giving rise to an unbounded class of new diffeomorphism-invariant observables. Such observables, which defined on codimension-one of bulk region, can be written as
	\begin{align}\label{obser}
	\mathcal{O}_{F_1,\Sigma_{F_2}}(\Sigma_{CFT})=\frac{1}{G_N L}\int_{\Sigma_{F_2}}\mathrm{d}^{d}\sigma \sqrt{h} F_1(g_{\mu\nu};X^{\mu})
	\end{align}
	where, $F_1$ is not necessarily equals to one, which led to the volume of hypersurface as in CV. Instead, it can be a general scalar function of bulk geometry, more precisely, metric $g_{\mu\nu}$  and an embedding $X^{\mu}(\sigma^a)$ of the hypersurfaces. Furthermore, $\Sigma_{F_2}$ is codimension-one hypersurface in the bulk spacetime with boundary time slice $\partial\Sigma_{F_2} =\Sigma_{CFT}$. On the other hand, extremality of the hypersurface leads to
	\begin{align}\label{var}
	\delta_{X}\Big(\int_{\Sigma}\mathrm{d}^{d}\sigma \sqrt{h} F_2(g_{\mu\nu};X^{\mu})\Big)=0. 
	\end{align}
	For simplicity, we follow the case $F_1 =F_2$, so the observable \eqref{obser}, denoted by generalized complexity $\mathcal{C}_{gen}$, obeys above condition, so
	\begin{align}\label{maxv}
	\mathcal{C}_{gen}(\tau)= \max_{\partial\Sigma(\tau)=\Sigma_{\tau}}\frac{V_x}{G_N L}\left[\int_{\Sigma}\mathrm{d}^{d}\sigma \sqrt{h} F_1(g_{\mu\nu};X^{\mu})\right]
	\end{align}
	where $h$ is the determinant of induced metric on the given hypersurface. In the original literature of CAny conjecture \cite{any,any?,anys} and other examples \cite{omidi,RN,gbcom,pht,mp,ks}, the authors specifically assume
	\begin{align}\label{F1F2}
		F_1=F_2=1+\lambda L^4 C^2,
	\end{align}
	where $C^2$ is the Weyl tensor squared and $\lambda=0$ corresponds to the CV prescription. There are also other assumptions like $ 1+\lambda L^4 C^2+\lambda_4 L^8 C^4$ \cite{anys}\footnote{In some of these literature $F_1\ne F_2$ is also perused \cite{any,pht}.}. In this paper, we are going to study the generalized complexity of Lovelock's theory as a generalization to Einstein gravity with higher curvatures. Our focus is on the second and third-order terms of the theory. In addition, in the case of Gauss-Bonnet gravity, we extend the assumption \eqref{F1F2} to include four invariant terms in $F_1$. \\	
	The remainder of this paper is organized as follows. In section \ref{sec: gencom}, we review the generalized complexity proposal. In section \ref{sec: lovelock}, we briefly review Lovelock theory. Then, we numerically calculate the time dependence of the generalized complexity, specifically, for the second-order term of the theory (Gauss-Bonnet) and the third one, respectively in subsections \ref{sec: GB} and \ref{sec: 3rd}. Section \ref{lt}, corresponds to the behavior of generalized complexity growth rate at late time. In subsection \ref{sec:therm}, we discuss the linear proportionality to the difference of the product of temperature and entropy at the two horizons, and we suggest that this can be corrected for the generalized case. We support these arguments with numerical analyses for the cases considered in section \ref{sec: lovelock}. In subsection \ref{sec:sing}, we study this behavior in the near singularity limit, which can be described by the Kasner metric. We then find possible values for complexity growth rate for Einstein-Hilbert and the first two Lovelock theories, both charged and uncharged, for different generalization functions. Finally, we conclude in section \ref{sec: conclusion}.  
	\section{Generalized volume complexity}\label{sec: gencom}
	
\begin{figure}[!htb]
		\centering
		\begin{tikzpicture}
			\coordinate (A) at (0,0);
			\coordinate[above = 2.25cm of A] (B);
			\coordinate[above = 2cm of B] (C);
			\coordinate[above = 2cm of C] (D);
			\coordinate[above = 2.25cm of D] (E);
			\coordinate[right = 2cm of A] (F);
			\coordinate[right = 4cm of C] (G);
			\coordinate[right = 2cm of E] (H);
			\coordinate[right = 4cm of A] (I);
			\coordinate[above = 2.25cm of I] (J);
			\coordinate[above = 2cm of J] (K);
			\coordinate[above = 2cm of K] (L);
			\coordinate[above = 2.25cm of L] (O);
			\coordinate[right = 0.25cm of F] (Q);
			\coordinate[left = 0.25cm of F] (P);
			\coordinate[right = 0.25cm of H] (S);
			\coordinate[left = 0.25cm of H] (R);
			\draw (C) -- (K);
			\draw[dashed] (F) -- (H);
			\draw[draw=gray, thick] (B) -- (L)
			node [below, sloped, pos=0.25 ] {$r_+$}
			node [below, sloped, pos=0.75 ] {$r_+$};
			\draw[draw=gray, thick] (J) -- (D)
			node [below, sloped, pos=0.25 ] {$r_+$}
			node [below, sloped, pos=0.75 ] {$r_+$}; 
			\draw[draw=gray,thick, dotted] (D) -- (S)
			node [above, sloped, pos=0.5 ] {$r_-$};
			\draw[draw=gray,thick, dotted] (L) -- (R)
			node [above, sloped, pos=0.5 ] {$r_-$}; 
			\draw[draw=gray,thick, dotted] (B) -- (Q)
			node [below, sloped, pos=0.5 ] {$r_-$};
			\draw[draw=gray,thick, dotted] (J) -- (P)
			node [below, sloped, pos=0.5 ] {$r_-$}; 
			\draw[draw, blue, very thick]  (0,4.95) arc (250:290:5.87cm)
			node[below, blue, pos=0.75 ] {$\Sigma$}  node[above, pos=0.5] {$r_{min}$};
			\draw[draw=red, dashed]  (4,2.25) arc (190:170:11.667cm);
			\draw[draw= red, dashed]  (0,2.25) arc (-10:10:11.667cm) node[ red, pos=0.3, below, sloped ] {$r_{max}$};
			\draw[ultra thick] (B) -- node[left, pos=0.5 ] {$r=\infty$} node[left, blue, pos=0.67 ] {$t_L=\tau /2$} node[left, green, pos=0.99 ] {$\tau=\infty$} (D);
			\draw[ultra thick] (J) -- node[right, pos=0.5 ] {$r=\infty$} node[right, blue, pos=0.67 ] {$t_R=\tau /2$} (L) node[right, green, pos=0.99 ] {$\tau=\infty$};
			\draw[draw=green, thick]  (0,6.3) arc (222:318:2.69cm) node[above, green, pos=0.5] {$r=r_{f}$};
			\draw[ultra thick, dotted] (A) -- node[left, pos=0.5 ] {$r=0$} (B);
			\draw[ultra thick, dotted] (D) -- node[left, pos=0.5 ] {$r=0$} (E);
			\draw[ultra thick, dotted] (I) -- node[right, pos=0.5 ] {$r=0$} (J);
			\draw[ultra thick, dotted] (L) -- node[right, pos=0.5 ] {$r=0$} (O);
		\end{tikzpicture}
		\caption{Penrose diagram of a two sided black hole with dual field theory in boundary. Dashed line in the middle shows minimum radius of spacetime. Red dashed curves on both sides near the border show maximum value of the radius. Blue curve is hypersurface in bulk theory anchored in the boundary time of both sides.The green curve shows the hypersurface maximizing effective potential and boundary time tends to infinity, and obviously it completely trap inside horizons.}\label{fig:1}
	\end{figure}
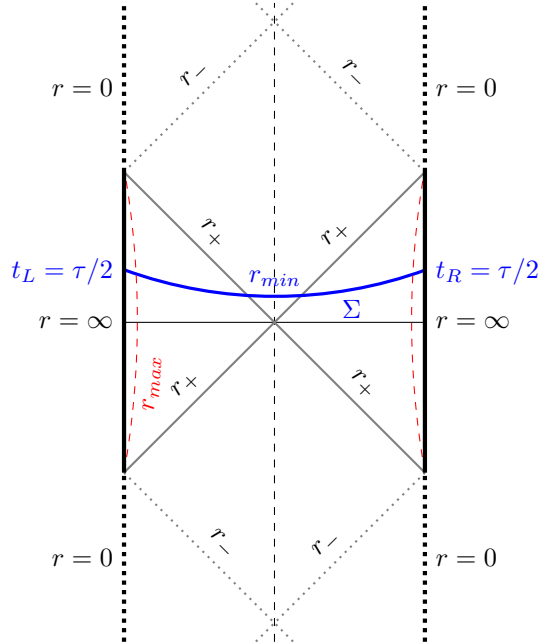

	Here, we review the generalized volume complexity. The set-up includes a two-sided black hole with the dual field theory in boundary. The corresponding Penrose diagram is depicted schematically in Fig. \ref{fig:1}. Boundary time is shown on both sides and their infinity. Dashed line in the middle shows minimum radius of spacetime that reaches zero in the top and bottom, which is the singularity in theory of gravity in bulk. Red dashed curves, on both sides near the border, show the maximum value of the radius. Also, the blue curve shows a hypersurface anchored at a specific time in the dual boundary theory, reaching time infinity limit in green curve. \\
	For the Eddington-Finkelstein coordinates, metric is written as
	\begin{align}\label{metricEF}
		\mathrm{d} s^{2}=-f(r) \mathrm{d} v^{2}+2 	\mathrm{~d}v\mathrm{~d}r+\big(\frac{ r}{L}\big)^2 \mathrm {d} \Omega _{k,d-1}^2, 
	\end{align}
	where
	\begin{align}
		v=t+r^{*}(r), \quad \mathrm{d}r^{*}=\frac{\mathrm{d}r}{f(r)},		
	\end{align}
	and $\mathrm {d}\Omega_k^2$ is $(d-1)$-dimensional metric which can take spherical, planar and hyperbolic geometry, respectively, with $k=1,0,-1$ as curvature.
	The generalized volume complexity as a codimension-one observable could be obtained from \eqref{maxv} and \eqref{F1F2} as \cite{any,any?,anys}
	\begin{align}\label{cgen}
		\mathcal{C}_{gen}(\tau)= \frac{V_0}{G_N L}\int_{\Sigma}\mathrm{d}\sigma \big(\frac{ r}{L}\big)^{d-1}\sqrt{-f \dot{v}^{2}+2 \dot{v} \dot{r}} a(r)
	\end{align}
	where dots indicate derivative with respect to $\sigma$, $V_0$ is volume of spatial spherical submanifold, and $a(r)$ is given by plugging the background (\ref{metricEF}) in $F_1$.
	Extremizing $\mathcal{C}_{gen}$ is equivalent to solving equations of motion when considering \eqref{cgen} as an action. Moreover, because spacetime is stationary, the momentum conjugate to coordinate $v$ is conserved
	\begin{align}
		P_v=-\frac{\partial \mathcal{L}}{\partial \dot{v}}=\frac{a(r)(r/L)^{d-1}(\dot{r}-f(r) \dot{v})}{\sqrt{-f \dot{v}^{2}+2 \dot{v} \dot{r}}}=\dot{r}-f(r) \dot{v}.	
	\end{align}
	Since expression in \eqref{cgen} is invariant under reparametrization, one can fix parameter $\sigma$ by choosing 
	\begin{align}
		\sqrt{-f \dot{v}^{2}+2 \dot{v} \dot{r}}=a(r)\big(\frac{ r}{L}\big)^{d-1}.
	\end{align}
	Then it is straightforward to derive extremality conditions with two equations above as follows
	\begin{align}\label{rdot}
		\dot{r}&=\pm \sqrt{P^2_v +f(r)a(r)^2\big(\frac{r}{L}\big)^{2(d-1)}},\\
		\label{tau}
		\dot{t}&=\dot{v}-\frac{\dot{r}}{f(r)}=\frac{-P_v\dot{r}}{f(r) \sqrt{P^2_v + f(r)a(r)^2(r/L)^{2(d-1)}}}.
	\end{align}
	We can see the equation as of classical particle's motion 
	\begin{align}
		\dot{r}^{2}+U(r)=P^2_v
	\end{align}
	with the effective potential as
	\begin{align}
		U(r)=-f(r)a(r)^2\big(\frac{r}{L}\big)^{2(d-1)}.
	\end{align}	
	The conserved momentum can be simply expressed as a function of turning point $r_{min}$ in the symmetric trajectory
	 \begin{align}
	 	P^2_v=U(r_{min})=-
	 	f(r_{min})a(r_{min})^2\big(\frac{r_{min}}{L}\big)^{2(d-1)}.
	 \end{align}	
	On the other hand, by integrating \eqref{tau}, the conserved momentum would be fixed in terms of boundary time, which is in turn a function of turning point $r_{min}$
	 \begin{align}\label{tauint}
		\tau & =2 \int_{r_{\min }}^{\infty} \mathrm{d} r \frac{-P_v}{f(r) \sqrt{P^2_v + f(r)a(r)^2(r/L)^{2(d-1)}}}.
	\end{align}
	As a result of symmetry, for coordinate $v$ in boundary and turning point we have
  	\begin{align}
    	v_{\infty}-v_{\min }=\tau_{\mathrm{R}}+r^{*}(\infty)-r^{*}\left(r_{\min }\right)
   	\end{align}
    Further, integrating extremality relation of coordinate $v$
   	\begin{align}
   		& v_{\infty}-v_{\min }=\int_{v_{\min }}^{v_{\infty}} \mathrm{d} v \nonumber\\
   		& =\int_{r_{\min }}^{\infty} \mathrm{d}r\left[\frac{P_v}{f(r) \sqrt{P^2_v + f(r)a(r)^2(r/L)^{2(d-1)}}}+\frac{1}{f(r)}\right] .
    \end{align}
	Hence, one can show that $\mathcal{C}_{gen}$ is  
	\begin{align}
		\mathcal{C}_{gen}=&\frac{2V_0}{G_N L}\Bigl\{\int_{r_{\min }}^{\infty} \mathrm{d}r\frac{1}{f(r)}\left[\sqrt{P^2_v + f(r)a(r)^2(r/L)^{2(d-1)}}+P_v\right]\nonumber \\
		&-P_v\left(\tau_{\mathrm{R}}+r^{*}(\infty)-r^{*}\left(r_{\min }\right)\right)\Bigr\}.
	\end{align}
	Derivation of $\mathcal{C}_{gen}$ with respect to boundary time leads to 
	\begin{align}\label{cdot}
		\frac{\mathrm{d} \mathcal{C}_{gen}}{\mathrm{~d} \tau}=\frac{1}{2} \frac{\mathrm{d} \mathcal{C}_{gen}}{\mathrm{~d} \tau_{\mathrm{R}}}&=\frac{V_0}{G_N L} P_v\nonumber\\
		&=\frac{V_0}{G L} \sqrt{-f(r_{min})}a(r_{min})\big(\frac{r_{min}}{L}\big)^{d-1}.
	\end{align}
	Now, we can study growth rate of generalized complexity by conserved momentum, which is related to the boundary time as both are functions of $r_{min}$. Then, the late-time behavior of growth rate is determined by 
	\begin{align}\label{tauinf}
		\lim _{\tau \rightarrow \infty} \frac{\mathrm{d} C_{gen}}{\mathrm{~d} \tau}=\frac{V_0}{G L} \sqrt{-f\left(r_f\right)} a(r_f)\big(\frac{r_f}{L}\big)^{d-1},
	\end{align}
	where $r_f$ is local maximum of the
	effective potential, at which time goes to infinity.\\
	So for, stationary solutions of gravitational theories, analyzing time versus conserved momentum by \eqref{tauint} gives full time behavior of  generalized complexity growth rate. Besides, for late time behavior replacing solutions of effective potential maxima in \eqref{tauinf} is straightforward.  
	\section{Lovelock theory}\label{sec: lovelock}
	
	In this section, we review Lovelock theory of gravity. The Lagrangian in D=d+1 dimensions with negative cosmological constant and contribution of Maxwell field is given by \cite{ll, ll2}
	\begin{align}\label{lag}
		\mathcal{L}=\frac{1}{16\pi G_N}\Big[R+\frac{d(d-1)}{L^2}+\sum_{n=2}^{[d/2]}\alpha_n \frac{(d-2n)!}{(d-2)!}(-1)^n L^{2n-2} \chi_{2n}\Big] - \frac{1}{4\pi} F_{\mu\nu}F^{\mu\nu}
	\end{align}
	where $L$ is radius of AdS, $\lambda_n$ are arbitrary coupling constants, and $\chi_{2n}$ are Euler densities made by generalized Kronecker delta and Riemann tensors
	\begin{align}
		\chi_{2n}=\frac{1}{2^n}\delta^{\mu_1... \mu_n}_{\nu_1... \nu_n}R^{\nu_1\nu_2}_{\mu_1\mu_2} ... R^{\nu_{2n-1}\nu_{2n}}_{\mu_{2n-1}\mu_{2n}},
	\end{align}		
	where 
	\begin{align}
		\delta^{\mu_1... \mu_n}_{\nu_1... 	\nu_n}=(2n)!\delta^{[\mu_1... \mu_n]}_{\nu_1... \nu_n}.
	\end{align}		
	The first and second terms in \eqref{lag} refer to Einstein gravity Lagrangian and negative cosmological constant 
	\begin{align}  
		\Lambda&=-\frac{d(d-1)}{L^2}
	\end{align} 
	respectively. The first two coupling constants of higher order terms
	\begin{align}\label{alpha} 
		\tilde{\alpha}=(d-2)(d-3)\alpha_2,\
		\tilde{\alpha}^2=72{d-2 \choose 4}\alpha_3.
	\end{align}  \\
	The static black holes solutions take the form \cite{time-com5}
	\begin{align}\label{metric}
		ds^2=-f(r)\mathrm{d}t^2+\frac{1}{f(r)}\mathrm {d}r^2+\big(\frac{ r}{L}\big)^2 \mathrm {d} \Omega _{k,d-1}^2, 
	\end{align}
	 We focus on spherical ($k=1$) black holes in this paper. We note that the above metric reads \eqref{metricEF} in Eddington-Finkelstein coordinates. \\
	In the following of this section, we focus on two simpler cases, second and third orders of theory and investigate their time dependence behavior of generalized complexity which is proportional to the conserved momentum.
	\subsection{AdS Gauss-Bonnet}\label{sec: GB}
	
	For the second order of Lovelock gravity we have Gauss-Bonnet (GB) term as \cite{time-com5}
	\begin{align}
		\mathcal{L}_{2}=\frac{1}{16\pi G_N}\Big[\alpha_2\Big(R^2-4R_{\mu\nu}R^{\mu\nu}+R_{\mu\nu\rho\sigma}R^{\mu\nu\rho\sigma}\Big) \Big] - \frac{1}{4\pi}F_{\mu\nu}F^{\mu\nu}
	\end{align}
	which has blackening function of the metric in form \eqref{metric} or \eqref{metricEF} for charged black hole as
	\begin{align}
		f(r)=1+\frac{r^2}{2\tilde{\alpha}}\Big(1- \sqrt{1+4\tilde{\alpha}\big(\frac{M}{r^d}-\frac{1}{L^2}-\frac{Q^2}{r^{2d-2}}\big)}\Big)\;,
	\end{align}
	with coupling constant as \eqref{alpha} and
	\begin{align}
		M \equiv \frac{16 \pi G \tilde{M}}{(d-1) \Omega_{d-1}}, \quad Q^{2} \equiv \frac{2(d-2) G \tilde{Q}^{2}}{(d-1)}.
	\end{align}	
	The ADM mass of the black hole in terms of inner and outer horizon radius $r_{\pm}$ - because of two horizons for charged black holes - and charge $\tilde{Q}$ would be
	\begin{align}
		\tilde{M}=\frac{(d-1) \Omega_{d-1} r_{\pm}^{d-2}}{16 \pi G}\left(1+\frac{\tilde{\alpha}}{r_{\pm}^{2}}+\frac{r_{\pm}^{2}}{L^{2}}+\frac{2(d-2) G \tilde{Q}^{2}}{(d-1) r_{\pm}^{2 d-4}}\right).
	\end{align}
	One can calculate the Hawking's temperature and Wald's entropy as
	\begin{align}
		T=\frac{f'(r)}{4\pi}\Big|_{r=r_{\pm}}= \frac{d r_{\pm}^4+(d-2)L^2 r_{\pm}^2+\tilde{\alpha}(d-4)L^2}{4 \pi L^{2}r_{\pm}(2\tilde{\alpha}+r_{\pm}^2)} -\frac{(d-2)^{2} G \tilde{Q}^{2} r_{\pm}^{5-2d}}{2 \pi(d-1)(2\tilde{\alpha}+r_{\pm}^2)},
	\end{align}
	\begin{align}
		S= \frac{\Omega_{d-1}r_{\pm}^{d-1}}{4 G}\Big(1+\frac{2(d-1)\tilde{\alpha}}{(d-3)r_{\pm}^2}\Big)
	\end{align}
	We consider simpler case, five dimensional $D=d+1=5$, so we get the square of Weyl tensor as
	\begin{align} \label{C2}
		C^2\equiv C_{\mu\nu\rho\sigma} C^{\mu\nu\rho\sigma}=\frac{\left(r^2 f''(r)-2 r f'(r)+2 f(r)-2\right)^2}{2 r^4}.
	\end{align}
	The Kretschmann scalar is  
	\begin{align}
		R_{\mu\nu\rho\sigma} R^{\mu\nu\rho\sigma}=f''(r)^2+\frac{6 f'(r)^2}{r^2}+\frac{12 (f(r)-1)^2}{r^4},
	\end{align}
	The square of Ricci tensor and Ricci scalar, which are non-constant for Lovelock spacetime geometry can be written respectively as
	\begin{align}
		R_{\mu\nu}R^{\mu\nu}=\frac{r^2(r f''(r)+3  f'(r))^2 + 6(r f'(r)+2 f(r)-2)^2}{2r^4},
	\end{align}
	\begin{align}\label{R}
		R=\frac{r^2 f''(r)+6 r f'(r)+6 f(r)-6}{r^2}.
	\end{align}	
	Now, by these scalar functions, we are going to build the generalization factor $a(r)$. We consider the following two cases:\\ 

\begin{figure}[!htb]
	\centering
	\subfloat[]{\includegraphics[width=6cm]{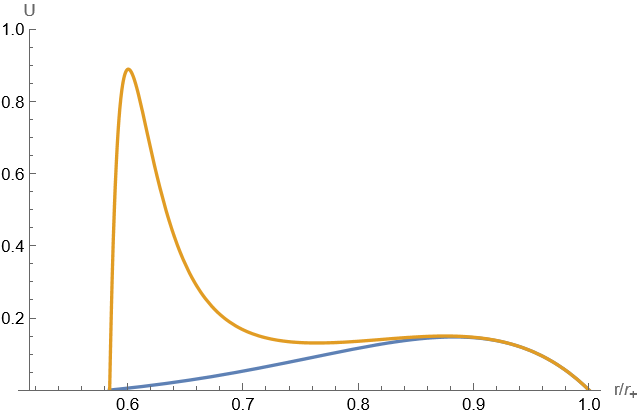}\label{fig:2a}}\qquad
	\subfloat[]{\includegraphics[width=6cm]{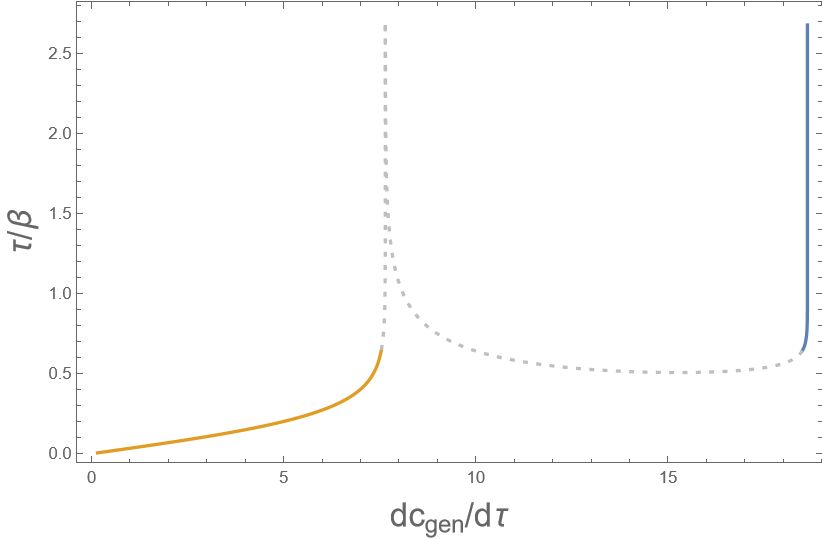}\label{fig:2b}}
	\subfloat[]{\includegraphics[width=6cm]{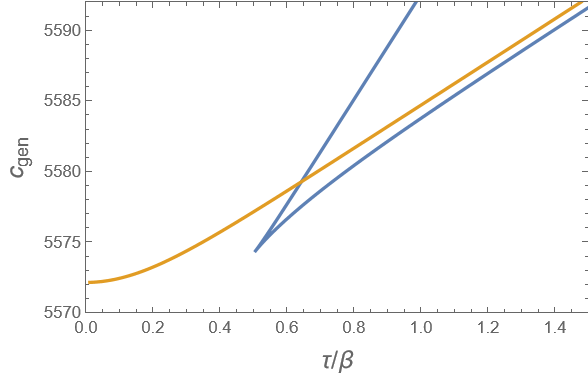}\label{fig:2c}}
	\caption{(a) Effective potential $U$ of GB in terms of dimensionless parameter $r/r_+$ (the blue curve corresponds to $\lambda=0$ $(\mathcal{C}_{V})$ while the orange to generalized), (b) boundary time multiplied by temperature $T=1/\beta$ vs. generalized complexity time rate and (c) generalized complexity vs. boundary time multiplied by temperature for the GB case of Weyl squared generalization term with $\tilde{\alpha}/r_+^2=0.2, L/r_+=1 $ and $M /r_+^2=3, \lambda=0.002$.}\label{fig:2}%
	\end{figure}

	\noindent $\bullet$ Case 1:	
	We first consider the Weyl squared based factor as
	\begin{align}\label{ac2}
		a(r)=a_1(r) \equiv 1+\lambda L^4 C^2,
	\end{align}
	the effective potential behavior in terms of dimensionless parameter $r/r_h$ for this case is shown in Fig. \ref{fig:2}\subref{fig:2a} in comparison to $\mathcal{C}_{V}$ i.e. when $a(r)=a_0(r) \equiv 1$. 
	From \eqref{tauinf}, we know that any local maxima of the effective potential with radius $r_f$ results in infinite boundary time. The conserved momentum at each $r_f$ is known as $p_{\infty}$. In this case, there are two peaks in comparison to one in $\mathcal{C}_{V}$.\\
	The numerical results of the boundary time versus  $\mathrm{d} C_{gen}/\mathrm{~d} \tau$, by \eqref{tauint} for \eqref{ac2} are shown in Fig. \ref{fig:2}\subref{fig:2b}. In all of calculations in this literature, $G=1$ is assumed.\\
	The time equation integral yields real solutions for regions where slope of the $\mathrm{d} C_{gen}/\mathrm{~d} \tau$ diagram is negative in terms of radius and is greater than the next maximum. The numbers of these regions clearly depends on local maxima or $p_{\infty}$ points, with each peaks in the diagram be smaller than the previous one.\\
	In the $\mathcal{C}_{V}$ case, there is only one descending region from the maximum to the end. In contrast, for $a_1(r)$ case, the effective potential have two peaks, and therefore two asymptotes appear in the $\tau-\mathrm{d} C_{gen}/\mathrm{~d} \tau$ diagram. The maximum points of momenta are denoted as $p_{\infty_{L}}$ and $p_{\infty_{R}}$, referring to the left and right sides of the diagram. There are three branches in the $\tau-\mathrm{d} C_{gen}/\mathrm{~d} \tau$ diagram, two of which approach $p_{\infty_{R}}$ from the left and right, while the third one approaches $p_{\infty_{L}}$ from the right.\\
	However, having different values of complexity rate at a single moment in time is not desirable. To determine the permissible intervals in each branch, another calculation for complexity as a function of boundary time is necessary.\\
	According to equation \eqref{cgen}, a similar numerical computation of complexity as a function of radius can be applied. Subsequently, the results can be plotted against boundary time as depicted in Fig. \ref{fig:2}\subref{fig:2c}. We observe three branches, where the single branch and upper one in V-shaped curve are dipping. As defined in maximum volume complexity (see \eqref{maxv}), intervals with higher values are considered acceptable. Notably, at certain points, there are intersections of curves where maximal volume exchange occurs from one branch to another. These points in boundary time are referred to as $\tau_{\text{turning}}$, signifying a phase transition of complexity and its corresponding time growth (or equivalently, conserved momentum). In Fig. \ref{fig:2}, the colors representing the complexity branches are consistent with those used to denote the allowed intervals before and after the phase transition in the $\tau-p$ diagram.\\
	
\begin{figure}[!htb]
		\captionsetup{width=0.8\textwidth}
		\begin{center}
		\subfloat[]{\includegraphics[width=6cm]{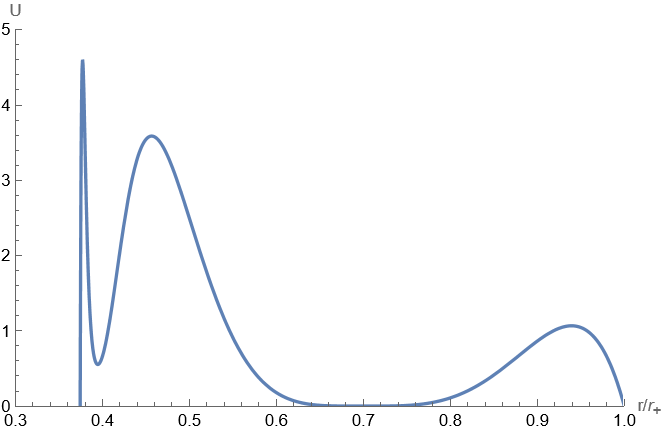} \label{fig:3a}}%
		\qquad
		\subfloat[]{\includegraphics[width=6cm]{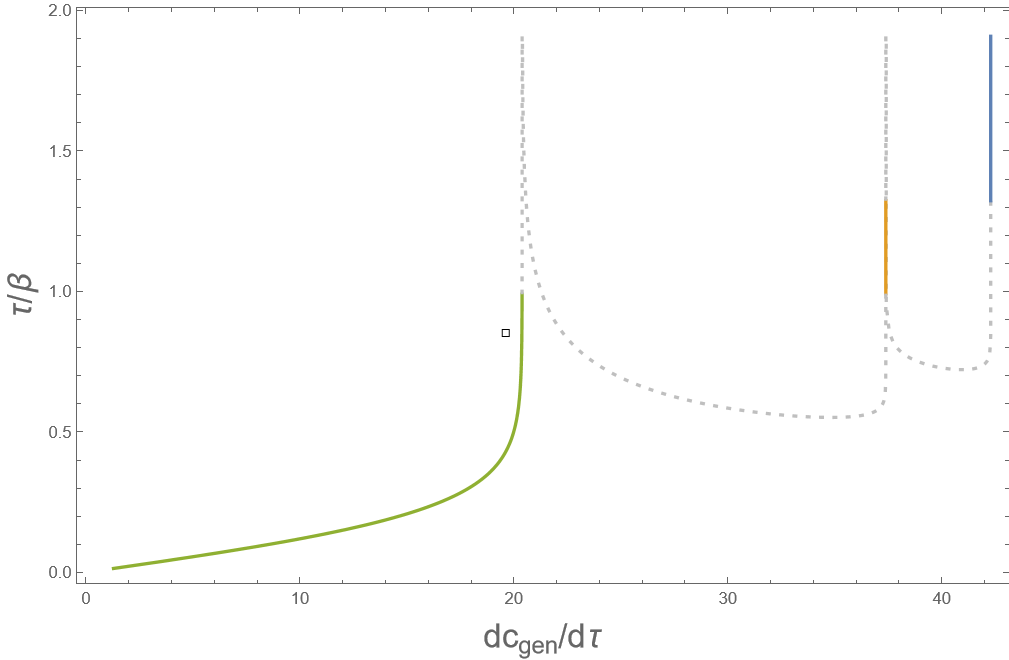} \label{fig:3b}}%
		\subfloat[]{\includegraphics[width=6cm]{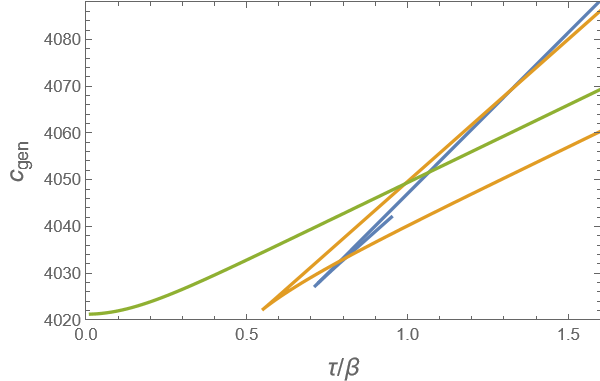} \label{fig:3c}}%
		\caption{ Effective potential $U$ of GB in terms of dimensionless parameter $r/r_+$, (b) boundary time multiplied by temperature $T=1/\beta$ vs. generalized complexity time rate and (c) generalized complexity vs. boundary time multiplied by temperature for the GB case of three generalization terms with $\tilde{\alpha}/r_+^2=0.2, L/r_+=1 $ and $M /r_+^2=2.5, \vec{\lambda}=(0.0001, -0.005, -0.008)$.}\label{fig:3}	
		\end{center}
	\end{figure}

	\noindent $\bullet$ Case 2:
	One can see that more additive term in $a(r)$ can result as extra maxima to the effective potential. For example we considered the case
	\begin{align}\label{ar2}
		a(r)=a_2(r) \equiv 1+\lambda_1 L^4 R_{\mu\nu\rho\sigma} R^{\mu\nu\rho\sigma}+ \lambda_2 L^4R_{\mu\nu}R^{\mu\nu}+ \lambda_3 L^4 R^2,
	\end{align}
	Such a factor can affect to potential as shown in Fig. \ref{fig:3}\subref{fig:3a}.\\
	The boundary time \eqref{tauint} versus the $\mathrm{d} C_{gen}/\mathrm{~d} \tau$ is depicted in Fig. \ref{fig:3}\subref{fig:3b}. There are three peaks, with each peaks in the diagram be smaller than the previous one.\\
	In this case with $a(r)=a_2(r)$, in addition to left and right peaks, we have an extra peak are corresponds to $p_{\infty_{M}}$ in the middle, resulting three branches in $\tau-\mathrm{d} C_{gen}/\mathrm{~d} \tau$ diagram, approach to three $p_{\infty}$'s from left, and two branches approach $p_{\infty_{M}}$ and $p_{\infty_{R}}$ from right. The first three are named as dipping branches \cite{any}.\\
	Similar to previous case, we performed a numerical computation of complexity versus the boundary time and depicted results in \ref{fig:3}\subref{fig:3c}. It can be seen that five branches show up from which the single branch and upper one in V-shaped curves are dipping. The phase transition points at $\tau_{\text{turning}}$ where curves intersect each others and maximal volume jumps from one branch to another one.  
	It is worth noting that for three peaks in the effective potential and five complexity branches, other types of phase transitions, one or none at all, are possible. Examples of such scenarios can be found in \cite{mp,pht}.
	\subsection{Third order}\label{sec: 3rd}
	
\begin{figure}[!htb]
		\centering
		\subfloat[]{\includegraphics[width=6cm]{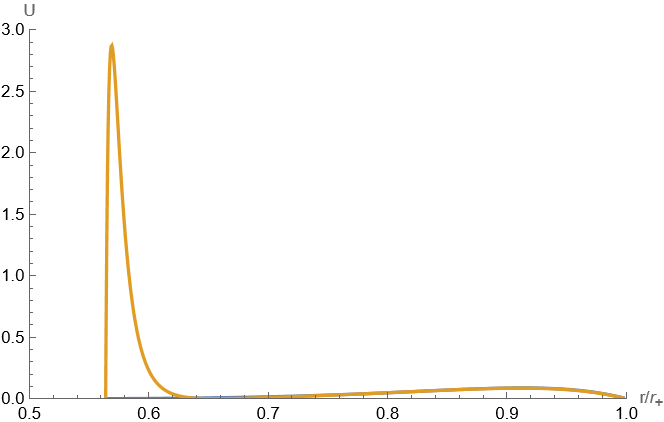} \label{fig:4a}}\qquad
		\subfloat[]{\includegraphics[width=6cm]{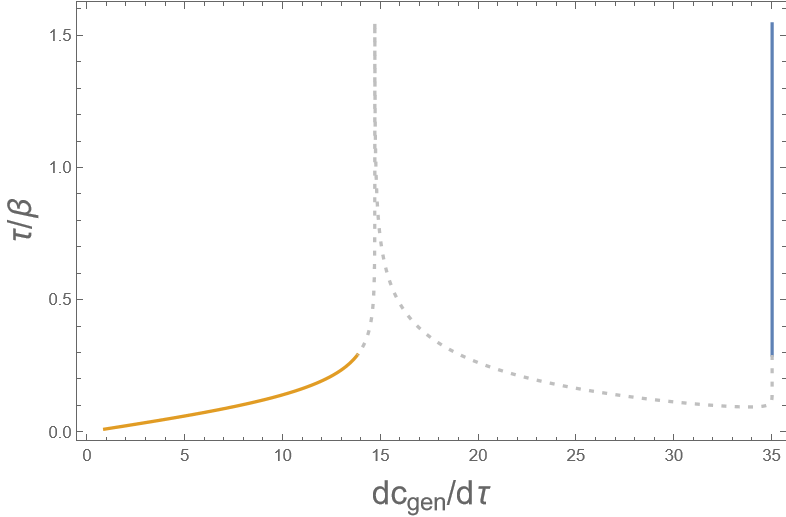} \label{fig:4b}}%
		\subfloat[]{\includegraphics[width=6cm]{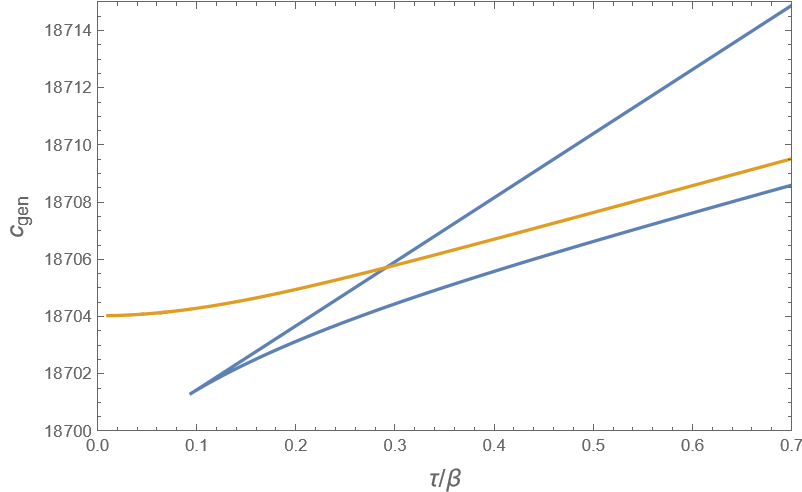} \label{fig:4c}}%
		\caption{(a) Effective potential $U$ of GB in terms of dimensionless parameter $r/r_+$ (the blue curve corresponds to $\lambda=0$ $(\mathcal{C}_{V})$ while the orange to generalized), (b) boundary time multiplied by temperature $T=1/\beta$ vs. generalized complexity time rate and (c) generalized complexity vs. boundary time multiplied by temperature for the GB case of Weyl squared generalization term with $\tilde{\alpha}/r_+^2=0.25, L/r_+=2 $ and $\tilde{M} /r_+^2=2, \lambda=-0.00005$.}\label{fig:4}%
	\end{figure}

	Taking one step forward, we consider the third order term of theory \cite{time-com4,3rd,ll2} with Lagrangian  
	\begin{align}
		\mathcal{L}_{3}&=\frac{1}{16\pi G_N}\Big[\alpha_3\Big(R^3+2R^{\mu \nu \sigma \kappa }R_{\sigma\kappa \rho \tau } R^{\rho \tau }_{\mu \nu }\nonumber \\&\quad +8R^{\mu \nu }_{\sigma \rho }R^{\sigma \kappa }_{\nu \tau } R^{\rho \tau }_{\mu \kappa }+24R^{\mu \nu \sigma \kappa }R_{\sigma \kappa \nu \rho } R^{\rho }_{\mu }\nonumber \\&\quad +3RR^{\mu \nu \sigma \kappa }R_{\mu \nu \sigma \kappa } +24R^{\mu \nu \sigma \kappa }R_{\sigma \mu }R_{\kappa \nu } \nonumber \\ {}&\quad +16R^{\mu \nu }R_{\nu \sigma }R^{\sigma }_{~\mu }-12RR^{\mu \nu }R_{\mu \nu }\Big)\Big] - \frac{1}{4\pi}F_{\mu\nu}F^{\mu\nu} 
	\end{align}\\
	and blackening function for slowly rotating charged solution we have 
	\begin{align}\label{bla}
		f(r)=1+\frac{r^2}{2\tilde{\alpha}}\biggl[1- \Big(1+6\tilde{\alpha}\big(\frac{M}{r^d}-\frac{1}{L^2}-\frac{Q^2}{r^{2d-2}}\big)\Big)^{1/3}\biggl]\;,
	\end{align}
	where we used \eqref{alpha} for coupling constant and again
	\begin{align}
		M \equiv \frac{16 \pi G \tilde{M}}{(d-1) \Omega_{d-1}}, \quad Q^{2} \equiv \frac{(d-2) G \tilde{Q}^{2}}{2(d-1)}.
	\end{align}	
	The ADM	mass in terms of $r_{\pm}$ and charge $\tilde{Q}$ would be
	\begin{align}
		\tilde{M}=\frac{(d-1) \Omega_{d-1} r_{\pm}^{d-2}}{16 \pi G}\left(1+\frac{2\tilde{\alpha}}{r_{\pm}^{2}}+\frac{4\tilde{\alpha}^2}{3r_{\pm}^{4}}+\frac{r_{\pm}^{2}}{L^{2}}+\frac{2(d-2) G \tilde{Q}^{2}}{(d-1) r_{\pm}^{2 d-4}}\right).
	\end{align}
One can calculate the Hawking's temperature and Wald's entropy as
	\begin{align}
		&T=\frac{f'(r)}{4\pi}\Big|_{r=r_{\pm}} \nonumber \\
		&=\frac{3d\tilde{\alpha}r_{\pm}^{2d+6}+L^2\big[3r_{\pm}^{2d+2}+r_{\pm}^{2d}\big(-3r_{\pm}^{6}+3(d-6)\tilde{\alpha}r_{\pm}^{4}+6(d-6)\tilde{\alpha}^2r_{\pm}^{2}+4(d-6)\tilde{\alpha}^{3}\big)-3(d-2)^{2} G \tilde{Q}^{2}r_{\pm}^{8}\big]}{12 \pi L^{2}\tilde{\alpha}r_{\pm}^{2d+1}(2\tilde{\alpha}+r_{\pm}^2)^2},
	\end{align}
	\begin{align}
		S= \frac{\Omega_{d-1}r_{\pm}^{d-1} }{4 G}\Big(1+\frac{2(d-1)\tilde{\alpha}}{(d-3)r_{\pm}^2}+\frac{(d-1)\tilde{\alpha}^2}{(d-5)r_{\pm}^4}\Big).
	\end{align}
	Here, we consider the case $a(r)=a_1(r)$ introduced in \eqref{ac2}.	
	For $D=d+1=7$, the lowest dimensions of spacetime in which this term contributes, we get square of Weyl tensor as
	\begin{align}
		C^2\equiv C_{\mu\nu\rho\sigma} C^{\mu\nu\rho\sigma}=\frac{2 \left(r^2 f''(r)-2 r f'(r)+2 f(r)-2\right)^2}{3 r^4}
	\end{align}
	Regarding \eqref{ac2}, the typical behavior of the effective potential in terms of dimensionless parameter $r/r_{\pm}$ for the third order Lovelock theory are shown in Fig. \ref{fig:4}\subref{fig:4a}. For the third order theory, similar to the GB case, adding Weyl squared term to $a(r)=1$, the number of maxima of the effective potential as a function of radius will increase to more than one. As discussed for GB, there are three branches for complexity growth rate due to two negative slope of potential, two of them approach to each $P_{\infty}$ from right, and the other approaches $P_{\infty_{R}}$ from left, which can be seen in boundary time versus the $\mathrm{d} C_{gen}/\mathrm{~d} \tau$ plot, Fig. \ref{fig:4}\subref{fig:4b}. Again, a third computation is done for complexity in time to find the phase transition point at $\tau_{\text{turning}}$. The corresponding diagram is depicted in Fig. \ref{fig:4}\subref{fig:4c}. 
	\\	
	\section{Late time behavior}\label{lt}
	  For complexity growth rate in late time, which means final size of black hole, due to spacial relation in different conjectures, i.e. proportionality to root of effective potential maximum in volume conjecture and its generalization, it is useful to study its behavior in this limit. In following subsections, we first deal with finding thermodynamics equivalent, then we explore near singularity by possible values of generalized complexity growth rate.
	\subsection{Thermodynamic quantities}\label{sec:therm}
\begin{figure}[!h]
		\centering
		\subfloat[]{\includegraphics[width=6cm]{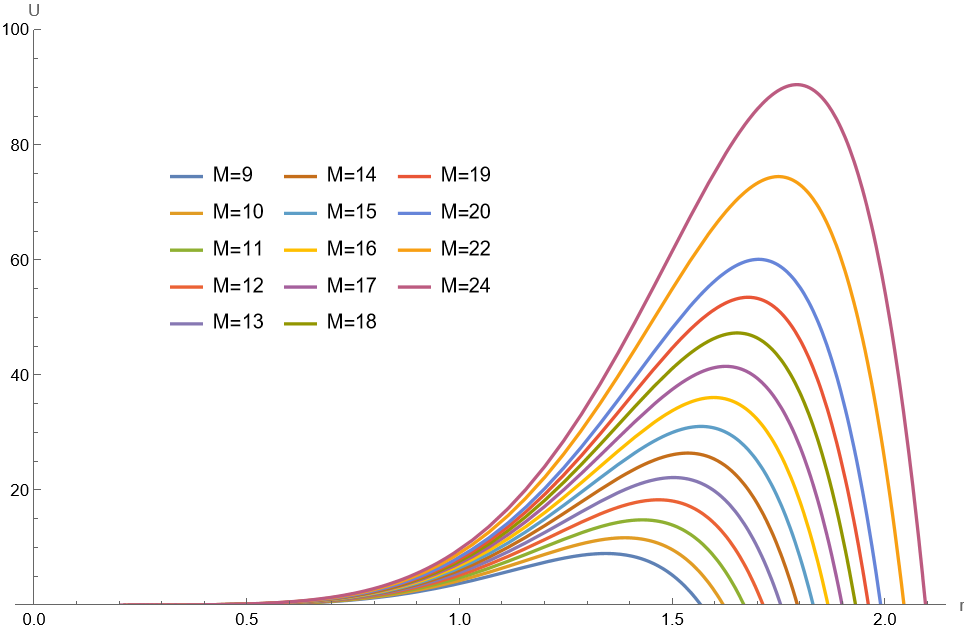} \label{fig:5a}}%
		\subfloat[]{\includegraphics[width=6cm]{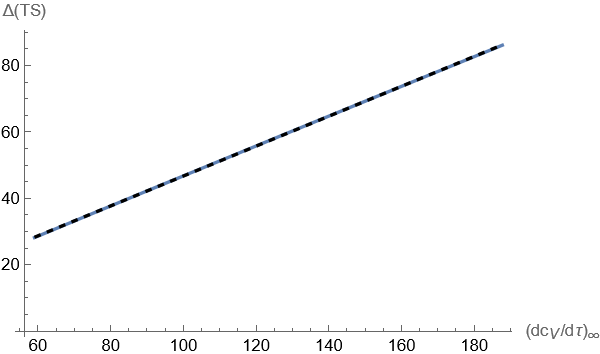} \label{fig:5b}}%
		\qquad
		\subfloat[]{\includegraphics[width=6cm]{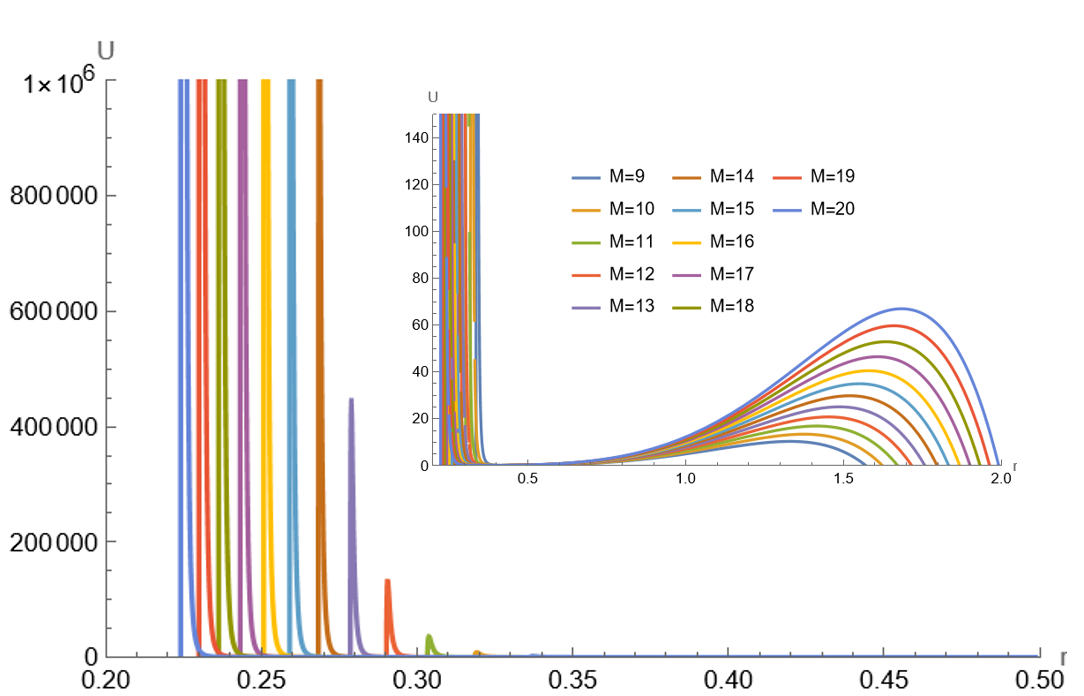} \label{fig:5c}}%
		\subfloat[]{\includegraphics[width=6cm]{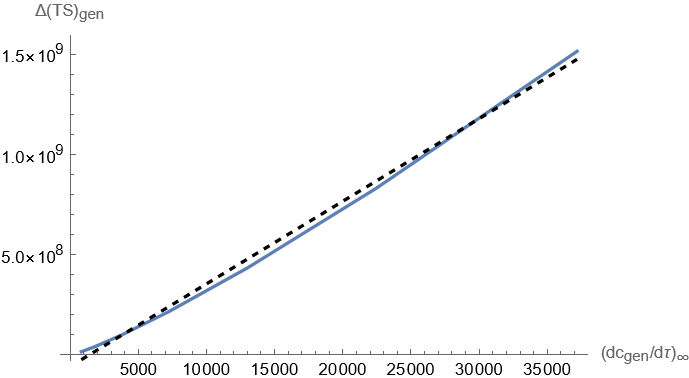} \label{fig:5d}}%
		\qquad
		\subfloat[]{\includegraphics[width=6cm]{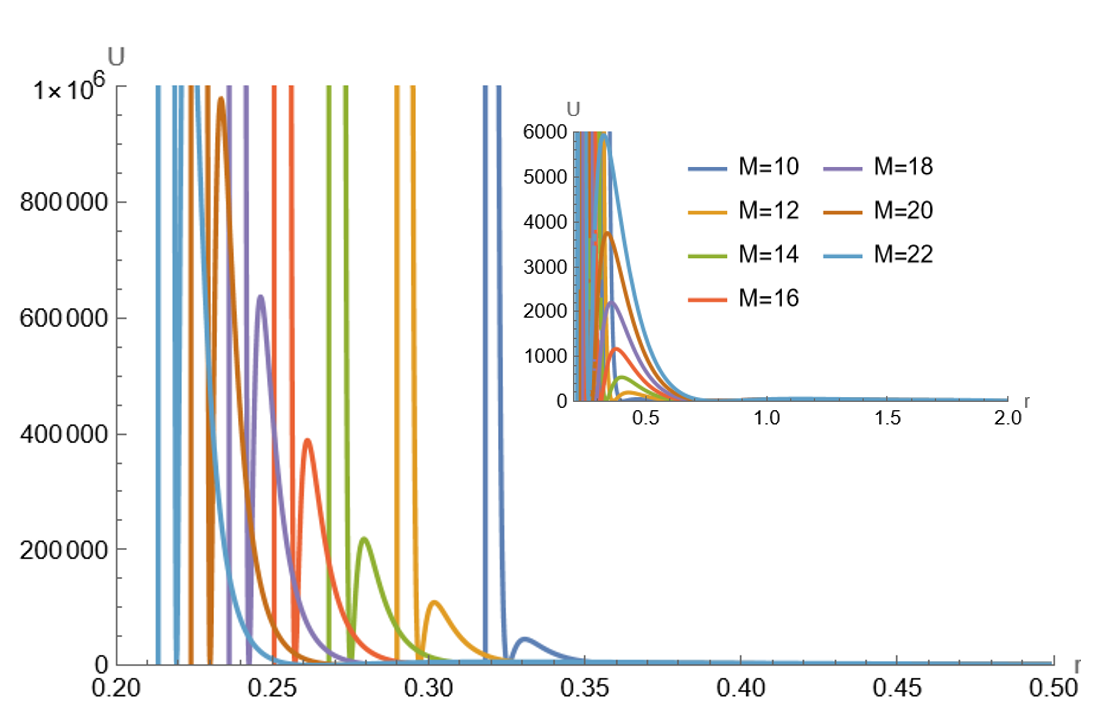} \label{fig:5e}}%
		\subfloat[]{\includegraphics[width=6cm]{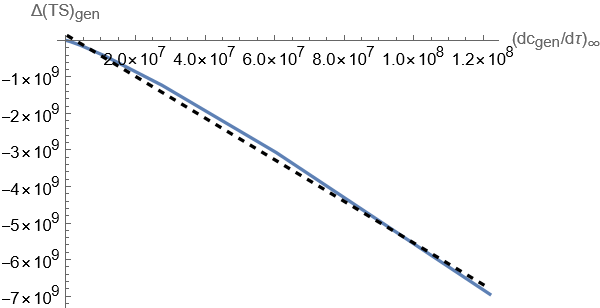} \label{fig:5f}}%
		\caption{Left: Effective potentials in GB theory with volume complexity ($a(r)=1$) and with generalization function of cases i and ii ($a(r)=a_1(r),a_2(r)$). Right: Proportionality of $\lim _{\tau \rightarrow \infty} \frac{\mathrm{d} C}{\mathrm{~d} \tau}$ to $\Delta(TS)_{gen}$ (blue line); Black dashed lines are linear fitting of blue lines. Parameters are fixed as $L=1, \alpha=0.1, Q=1$ (a,b) $\lambda=0$ (c,d) $\lambda=0.000005$ (e,f) $\vec{\lambda}=(0.0005, -0.0001, -0.001)$ , with different masses.}\label{fig:5}%
	\end{figure}

\begin{figure}[!h]
		\centering
		\subfloat[]{\includegraphics[width=6cm]{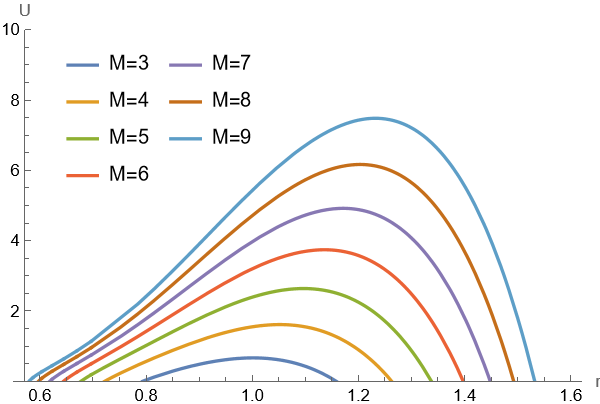} \label{fig:6a}}%
		\subfloat[]{\includegraphics[width=6cm]{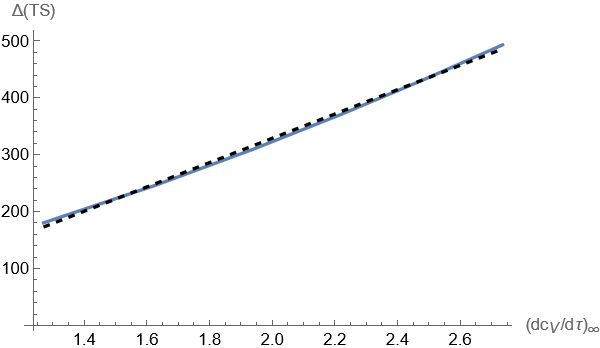} \label{fig:6b}}%
		\qquad
		\subfloat[]{\includegraphics[width=6cm]{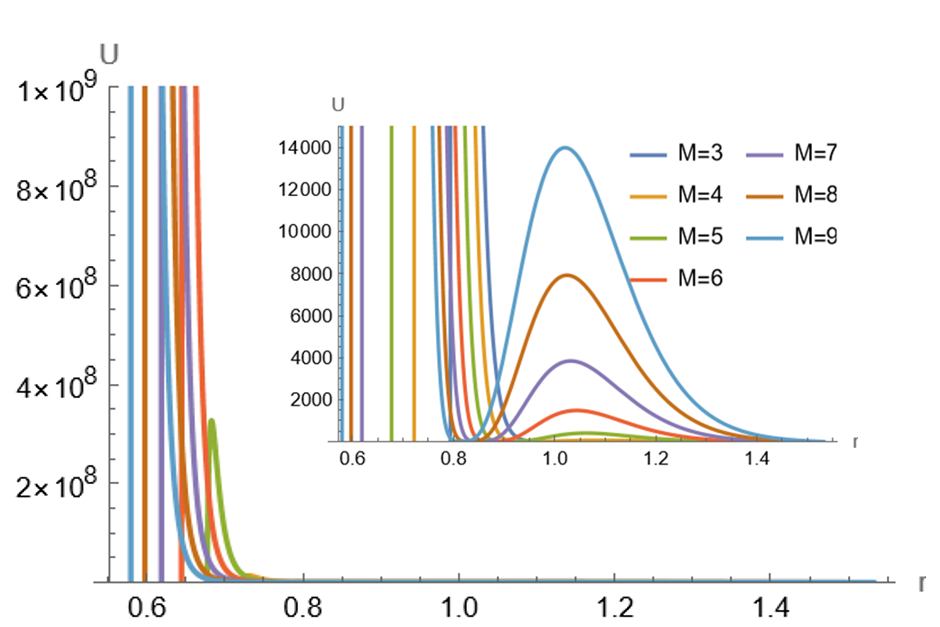} \label{fig:6c}}%
		\subfloat[]{\includegraphics[width=6cm]{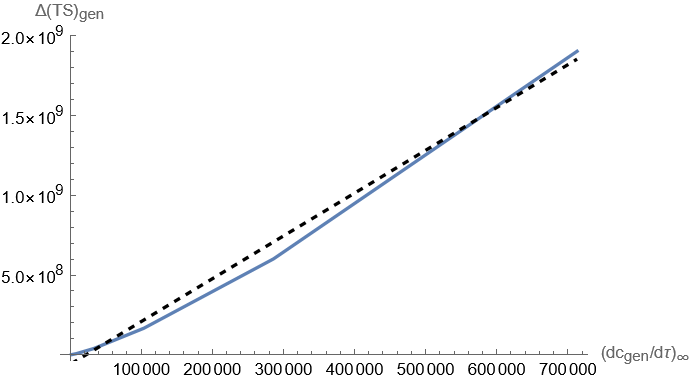} \label{fig:6d}}%
		\caption{Left: Effective potentials in GB theory with volume complexity ($a(r)=1$) and with generalization function case i ($a(r)=a_1(r)$). Right: Proportionality of $\lim _{\tau \rightarrow \infty} \frac{\mathrm{d} C}{\mathrm{~d} \tau}$ to $\Delta(TS)_{gen}$ (blue line); Black dashed lines are linear fitting of blue lines. Parameters are fixed as $L=2, \alpha=0.015, Q=1$ (a,b) $\lambda=0$ (c,d) $\lambda=0.005$, with different masses.}\label{fig:6}%
	\end{figure}

	As demonstrated in various literature, the growth rate of complexity in late times is directly proportional to temperature multiplied by entropy $TS$ of the black hole (e.g. \cite{time-com5, time-com3, ts1, ts2, ts3, ts4}) 
	\begin{align}
	&\lim _{\tau \rightarrow \infty} \frac{\mathrm{d} C_{V}}{\mathrm{~d} \tau}\sim TS
	\end{align}
	For the black holes without charge and rotation, with one horizon, it equals to the ADM mass. For charged or rotating black holes which have two horizons, complexity growth rate usually is proportional to difference of $TS$ in outer and inner horizons, 
	\begin{align}\label{TS}
	&\lim _{\tau \rightarrow \infty} \frac{\mathrm{d} C_{V}}{\mathrm{~d} \tau}\sim(TS)_+-(TS)_-\equiv\Delta(TS)
	\end{align}	
	Moreover, in \cite{any?}, preserving proportionality to ADM mass, and in \cite{any,ks}, to $TS$, are shown for the generalized complexity.	Here, we surmise correction of this feature for charged black holes for the volume and generalized complexity. To upgrade the proportionality relation (\ref{TS}) to the generalized complexity, it is expected that the right-hand side must be modified in some sort. 
	We suggest that for generalized complexity at late time, proportionality to difference of $TS$ in outer and inner horizons can be corrected by multiplying generalization function in each horizon
	\begin{align}\label{TSgen}
		&\lim _{\tau \rightarrow \infty} \frac{\mathrm{d} C_{gen}}{\mathrm{~d} \tau}\sim(TS)_+a(r_+)-(TS)_-a(r_-)\equiv\Delta(TS)_{gen}
	\end{align}
	To test this suggestion, we performed the computations for three examples: i) GB with generalized complexity with Weyl tensor squared, ii) GB with three-term generalization functions, and iii) third order Lovelock with Weyl tensor squared. 
	The numerical results are shown in Fig. \ref{fig:5} for GB in volume complexity cases i and ii, and in Fig. \ref{fig:6} for third order in volume complexity case iii. The left panel plots show the effective potentials of each case for different ADM masses, which can be directly calculated by their largest maxima complexity growth rate in late time. It is evident that in all cases for volume complexity (i.e. $a(r)=1$) late time values are linear with the difference of $TS$ in outer and inner horizons as \eqref{TS}, and for generalized cases, the numerical results are approximately in agreement with the modified linearity relation in \eqref{TSgen}. Due to two negative coefficients of generalization function in the (ii), the values of $a(r_{\pm})$ and $\Delta(TS)_{gen}$ are negative (\ref{fig:5}\subref{fig:5f}). Descending slope of diagram represents that constant factor depends on these coefficients, which insures positive complexity growth rate at late time as shown in \ref{fig:5}\subref{fig:5e}. 
	\\
	\\	
	\subsection{Near singularity limit}\label{sec:sing}
	As discussed by \cite{ks,eon1, eon2, eon3}, in the vicinity of singularity, spacetime is anisotropic, therefore it can be defined by Kasner black hole. Namely, for the metrics of form
	\begin{align}\label{metric2}
		ds^2=-f(r)\mathrm{d}t^2+\frac{1}{f(r)}\mathrm {d}r^2+g(r)\mathrm {d} \Omega _{d-1}^2, 
	\end{align}
	near the singularity, approximately
	\begin{align}\label{vic}
		f(r)=-b_n r^{-z},
	\end{align} 
	where $z$ depends on the model and for two cases of $Q=0$ and $Q\neq 0$ are given in Table \ref{table:1}. Function $g(r)$ is as before
	\begin{align}
		g(r)=\left(\frac{r}{L}\right)^2.
	\end{align}
	Then in this limit, the metric can be deformed to Kasner 
	\begin{align}\label{kasner}
		ds^2=-\mathrm{d}\tilde{r}^2+k_1 \tilde{r}^{2p_1}\mathrm{d}t^2+k_j \tilde{r}^{2p_{j}}\mathrm {d} \Omega _{d-1}^2, 
	\end{align}
	by transformation as
	\begin{align}\label{trans}
		\mathrm{d}\tilde{r}=\frac{\mathrm{d}r}{\sqrt{-f(r)}}.
	\end{align}
	It leads to
	\begin{align}\label{trans2}
		&f(r)=-k_1 \tilde{r}^{2p_1},
		&g(r)=k_j \tilde{r}^{2p_j}
	\end{align}
	with
	\begin{align}\label{}
		&\begin{cases}
			p_1=-\frac{z}{z+2}\\\\
			k_1=\Big(\frac{z+2}{2}\Big)^{\frac{-2z}{z+2}} \big(b_n\big)^{\frac{2}{z+2}}
		\end{cases}
		&\begin{cases}
			p_{\Omega}=p_j=\frac{2}{z+2}\\\\
			k_{\Omega}=k_j=\frac{1}{L^2}\Big(\frac{z+2}{2}\Big)^{\frac{4}{z+2}} \big(b_n\big)^{\frac{2}{z+2}}&(j=2,3,\ldots,d).
		\end{cases}
	\end{align}
	For black holes without charge ($Q=0$) and charged ones ($Q\neq0$), the dominant terms change. Therefore, $z$, $b_n$ in \eqref{vic}, the sum of exponents, and the sum of their squares are calculated and listed in Table \ref{table:1}. For uncharged black holes, the first exponent depends on the order of theory and the dimension. 
		\begin{center} 
		\begin{table}
			\centering
			\caption{Kasner parameters for charged and uncharged black holes}\label{table:1}
			\renewcommand\arraystretch{2}
			\begin{tabular}{|*{3}{c|}}
				\hline
				&{$Q=0$} & {$Q\neq0$} \\ \hline
				$z$&$\frac{d-2n}{n}$&$\frac{2(d-n-1)}{n}$\\ \hline
				$b_n$&$\frac{\big(2n\tilde{\alpha}M\big)^{1/n}}{2\tilde{\alpha}}$&$\frac{\big(-2n\tilde{\alpha}Q^2\big)^{1/n}}{2\tilde{\alpha}}$ \\ \hline
				$p_1$&$\frac{2n-d}{d}$&$\frac{1+n-d}{d-1}$\\ \hline
				$p_\Omega$&$\frac{2n}{d}$&$\frac{n}{d-1}$\\ \hline
				$k_1$&$\big(\frac{d}{2n}\big)^{\frac{(4n-2d)}{d}}\big(\frac{(2n\tilde{\alpha}M)^{\frac{1}{n}}}{2\tilde{\alpha}}\big)^{\frac{2n}{d}}$&$\big(\frac{d-1}{n}\big)^{\frac{2(n-d+1)}{d-1}}\big(\frac{(-2n\tilde{\alpha}Q^2)^{\frac{1}{n}}}{2\tilde{\alpha}}\big)^{\frac{n}{d-1}}$ \\ \hline
				$k_\Omega$&$\frac{1}{L^2}\left(\frac{d^2(2n\tilde{\alpha}M)^{\frac{1}{n}}}{8n^2\tilde{\alpha}}\right)^{\frac{2n}{d}}$&$\frac{1}{L^2}\left(\frac{(d-1)^2(-2n\tilde{\alpha}Q^2)^{\frac{1}{n}}}{2n^2\tilde{\alpha}}\right)^{\frac{n}{d-1}}$\\ \hline
				$\sum_{i=1}^{d}p_i$&$2n-1$&$\frac{d(n-1)+1}{d-1}$\\ \hline
				$\sum_{i=1}^{d}p_i^2$&$1+\frac{4n(n-1)}{d}$&$\frac{d^2+d(n^2-2n-2)+2n+1}{d^2-2d+1}$\\ \hline
				$p_c$&$2n-1+p_a$&$\frac{dn-(d-1)(1-p_a)}{d-1}$\\ \hline
			\end{tabular}
		\end{table}
	\end{center}	
	
\begin{figure}[!h]
		\centering
		\subfloat[]{\includegraphics[width=6cm]{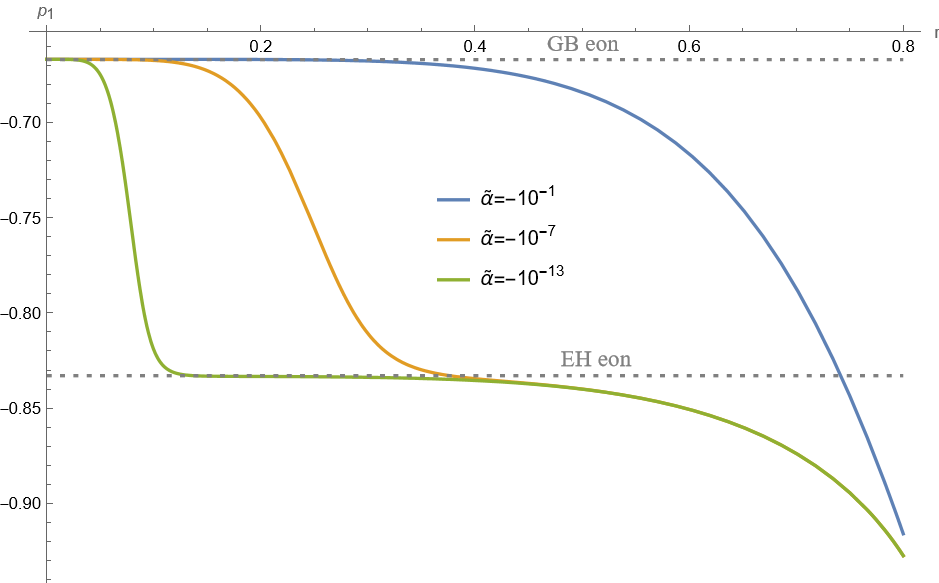} \label{fig:7a}}%
		\subfloat[]{\includegraphics[width=6cm]{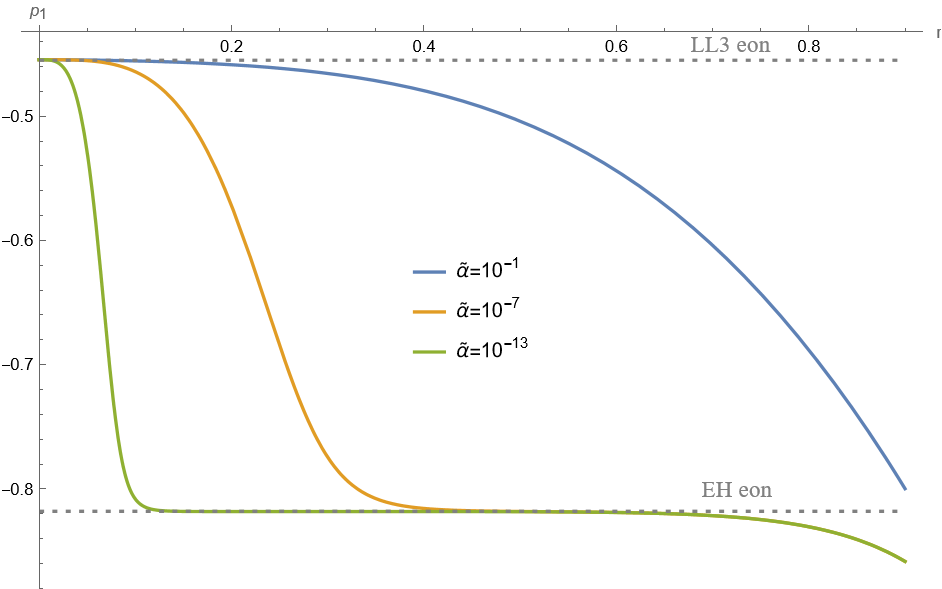} \label{fig:7b}}%
		\qquad 
		\caption{Stair-like eons of $p_1$ Kasner exponent at small radii in different coupling constants, for charged (a) GB with $d=7$ and (b) Lovelock third order in $d=11$.}\label{fig:7}%
	\end{figure}

	Plotting $P_1$ in terms of radius produces stair-like diagrams at small coupling constants. These diagrams show that as the singularity is approached, the term corresponding to the highest order becomes dominant, while for increasing radii, terms of lower orders prevail. 
	In \cite{eon1, eon2, eon3}, these diagrams are plotted for uncharged higher order theories, with these dominant ranges referred to as eons. Fig. \ref{fig:7} presents similar diagrams for our focus cases: charged second and third orders of Lovelock black hole solutions. We plotted $p_1$ Kasner exponent at small radii for charged theories, similar to uncharged ones in \cite{eon1, eon2, eon3}. It shows the Kasner eons stair-like diagrams. Fig. \ref{fig:7}\subref*{fig:7a}, corresponds to Gauss-Bonnet in $d=7$, and \ref{fig:7}\subref*{fig:7b} to Lovelock third order in $d=11$. In both cases for smaller coupling constants, approach to Einstein-Hilbert eon as getting away from singularity, and to Lovelock theory eon as approaching to it.
	\\	
	
	Here we study the behavior of the complexity depending on a generic generalization function $a(r)$. Define the near singularity limit as
	\begin{align}\label{}
	a(r)\equiv k_a \tilde{r}^{p_a},
	\end{align}
	Then from \eqref{tauinf}, we have
	\begin{align}\label{cgene}
		\lim _{\tau \rightarrow \infty} \frac{\mathrm{d} 		C_{gen}}{\mathrm{~d} \tau}=\frac{V_0}{G L} k_a k_1^{\frac{1}{2}} k_\Omega^{\frac{d-1}{2}}\tilde{r}^{p_c}.
	\end{align}
	where $p_c\equiv p_1+(d-1)p_{\Omega}+p_a$. The equation \eqref{cgene} indicates that certain constraints should be applied to the generalization function to annul the effect of $\tilde{r} \rightarrow 0$. Thus 
	values of late time generalized complexity growth rate for different choices of $a(r)$ become
	\begin{equation}\label{}
		\lim _{\tau \rightarrow \infty}\frac{\mathrm{d}C_{gen}}{\mathrm{~d} \tau}=\begin{cases}
			0, &\text{if $p_a>-\sum_{i=1}^{d}p_i$}\\\\
			\frac{V_0}{G L} k_a k_1^{\frac{1}{2}} k_\Omega^{\frac{d-1}{2}}, &\text{if $p_a=-\sum_{i=1}^{d}p_i$}\\\\
			\infty, &\text{if $p_a<-\sum_{i=1}^{d}p_i$}.
		\end{cases}
	\end{equation}
	If the second condition (i.e. $p_a=-\sum_{i=1}^{d}p_i$ or $p_c=0$) met, the finite nonzero generalized complexity growth rate can be found in the absence of charge ($Q=0$),
	\begin{align}\label{}
		&\lim _{\tau \rightarrow \infty}\frac{\mathrm{d}C_{gen}}{\mathrm{~d} \tau}=\frac{V_0}{G L^d} k_a \big(\frac{d}{2n}\big)^{2n-1}\frac{2n\tilde{\alpha}M}{(2\tilde{\alpha})^n},
	\end{align}
	which is proportional to $M$ as expected, and in the presence of charge $Q\neq0$
	\begin{align}\label{}
		&\lim _{\tau \rightarrow \infty}\frac{\mathrm{d}C_{gen}}{\mathrm{~d} \tau}=\frac{V_0}{G L^d}k_a\big(\frac{d-1}{n}\big)^{\frac{nd-d+n+1}{d}}
		\Big(\frac{-2n\tilde{\alpha}Q^2}{(2\tilde{\alpha})^n}\Big)^{\frac{d}{2d-2}}.
	\end{align} 
	In the presence of charge, for the odd orders $n$ theories, that $b_n$ is negative, due to power of $k_1$ and $k_\Omega$, only for spacetimes with even dimensions generalized complexity growth rate in late times has defined near singularity. In even orders $n$ theories, it is defined for more limited cases with negative coupling constants of higher order action terms, and spacetimes with integer multiples of $2n$ dimensions.
	
	Using (\ref{C2})-(\ref{R}), one can show that $p_a=-4$. If instead of (\ref{C2})-(\ref{R}), we choose higher order curvature terms for $a(r)$, we find $p_a=-2m$ where $m$ is the order of curvatures appear in $a(r)$. Then cases which have finite nonzero generalized complexity growth rate value near singularity constraint by the relation $p_c=-2m+\sum_i p_i=0$ and the fact that $d$, $n$ and $m$ are integers. For $m=2$ as given in (\ref{C2})-(\ref{R}), the only case is $d=5$ and $n=4$.  
	
	
	\section{Conclusion}\label{sec: conclusion}
	
	In summary, the study presented in this paper explored the time dependence of generalized complexity in the context of Lovelock gravity, which extends Einstein's theory by incorporating higher curvature terms. Specifically, we focused on two terms of the theory, the second order (GB) in 5 dimensions and the third one in 7. Our investigation centered on the time evolution of time growth rate of codimension-one generalized complexity ($\mathrm{d} C_{\text{gen}}/\mathrm{d}\tau$) via conserved momentum conjugate to the $v$ coordinate in Eddington-Finkelstein coordinates, under the condition of equal $F_1$ and $F_2$.\\		
	We used the generalization functions \eqref{ac2} and \eqref{ar2}. For the first term, associated with the $\mathcal{C}_{V}$ mode, a peak emerges in the effective potential or, similarly, conserved momentum diagram over the dimensionless radial quantity $r/r_+$. Adding other terms led to the appearance of two or three peaks, referred to as $P_{\infty}$. The numerical analysis of boundary time in terms of complexity growth rate revealed asymptotes at these points, and we studied the complexity against boundary time to determine the allowed branches. The maximal generalized volumes, defining the generalized complexity, were identified as the highest value intervals in these diagrams. Additionally, by computing the intersection points of different generalized volume curves, we identified times at which one maximal curve superseded another (i.e., $\tau_{\text{turning}}$). Consequently, we observed certain parts of dipping branches approaching late-time asymptotes from the right, elucidating how phase transitions occur at $\tau_{\text{turning}}$. These phase transitions correspond to a discontinuous jumps in the extremal surface \cite{gbcom}.\\ 
	Due to the shifting late time branch of the growth rate in this conjecture of complexity, new behaviors can emerge. One notable behavior is the linear relationship with the difference between $TS$ in outer and inner horizons for volume complexity growth rate of charged Lovelock black hole. For black holes with a single horizon, it is straightforward to show proportionality with $TS$ or $ADM$ mass. However, for those with two horizons, it is not as straightforward. We suggested that multiplying a generalized function of each horizon by these thermodynamic quantities serves as a correction for our generalized complexity in the case of charged Lovelock black holes. The subsequent numerical results confirmed a good approximation. Additionally, we deduced from the results that this proportionality should be a function of the coupling constants of the generalization terms. These correction suggestions can later lead to finding exact duals via thermodynamics in boundary theory for a new class of observables defined by generalized complexity. \\
	Another point of interest at late times is near the singularity limits defined by the Kasner regime. In higher dimensional theories of gravity, Kasner exponents and other parameters are related to their dimensions and orders, $(d,n)$ (see Table \ref{table:1}). The linear and quadratic sums of Kasner exponents in absence of charge ($Q=0$) are $2n-1$ and $(d+4n(n-1))/d$ which are well-known results (see e.g. \cite{eon2, eon3}). In a charged black hole ($Q\neq 0$), we derived these sums depending on both $d$ and $n$ as given in Table \ref{table:1}.	
	Consequently, we expect that dimension and order of theory play a crucial role in determining the exact values of the generalized complexity rate. It was discussed that different choices of the generalization function yield possible values of zero, finite, or infinity, depending on the dimensions and orders, for both charged and uncharged solutions of our relevant theory. These values are valid within the ranges governed by their respective orders of theory, referred to as eons. We added stair-like eon diagrams for charged black holes to those for uncharged black holes, which are similar to the results of \cite{eon3}. It is worth mentioning that in charged black holes, the term with a negative sign plays a crucial role in determining the Kasner parameters and complexity. Subsequently, we identified some constraints on the dimensions of spacetime and coupling constants. Indeed, the constraints on having finite value for the generalized complexity rate in late times are very restrictive, however it is possible to set parameters accordingly. For example, the finite rate can be achieved in $d=5$ and $n=4$ theory with a quadratic generalized function $a(r)$. 
	
	\section*{Acknowledgment}
	We would like to express our appreciation to Xuanha Wang and Yu-Xiao Liu for their invaluable guidance on the conceptual aspects of this study. We extend our sincere gratitude to Robie A. Hennigar for his invaluable comments and suggestions, which helped us enrich the quality of this paper. We also thank Ghadir Jafari and Shan-Ming Ruan for providing technical support. Additionally, we are grateful to Reza Fareghbal, Ali Naseh, and Mojtaba Shahbazi for their constructive comments.
	
   
\end{document}